\documentclass[aps,twocolumn,prl,showpacs]{revtex4}
  \usepackage[dvips]{color}
\usepackage{graphicx}
\usepackage{amsmath}
\usepackage[bf,Large]{}

\begin{document}

\title{\bf Reply to Comment on Circular Dichroism in the Angle-Resolved 
Photoemission Spectrum of the High-Temperature Bi$_2$Sr$_2$CaCu$_2$O$_8$ 
Superconductor, http://arxiv.org/abs/1004.1648}
\author{Matti Lindroos$^{1,2}$}\email{matti.lindroos@tut.fi}

\author{Ville Arpiainen$^{1}$}
\author{A. Bansil$^{2}$}
\affiliation{$^{1}$Institute of Physics,
Tampere University of Technology, P.O. Box 692, 33101 Tampere,
Finland}
\affiliation{$^{2}$Physics Department, Northeastern University, Boston,
Massachusetts 02115}


\maketitle

Our purpose in Ref. 1 was to carry out a realistic first-principles 
angle-resolved photointensity (ARPES) computation to assess the importance 
of geometric effects in inducing dichroism in the curpates, and not so 
much to focus on the results of Ref. 2. Using the known low- and 
high-temperature (LT and HT) bulk crystal structures of Bi2212, we showed 
that the orthorohombic distortion in the Bi-O plane in going from the HT 
to the LT structure yields an increase in the dichroic signal of 1.5 to 
3.5\%, depending on photon energy, which is comparable to the dichroic 
effect of 3\% reported in Ref. 2. Moreover, our computed dichroic signal 
of 6\% at 49.7 eV photon energy for the HT structure is comparable to the 
5\% effect reported on single crystals [3]. Thus, we clearly established 
the sensitivity of dichroism to structural details and the viability of 
the geometric mechanism in explaining existing measurements, and 
demonstrated that the detection of time reversal symmetry breaking via 
ARPES will be complicated by the masking effects of lattice distortions.

The structure of thin films of Ref. 2, especially how this structure 
varies with temperature/doping, is unknown. That the films were heavily 
twinned is revealed by the authors for the first time in their present 
Comment, but how the population of the two twins varies with doping or 
temperature is not known. Since the films display weak superstructure, one 
may speculate that these films are structurally closer to Pb-doped rather 
than non-Pb-doped crystals. Unfortunately, however, experiments 
on Pb-doped crystals [3] found a zero dichroic signal at low- as well as 
high-T, suggesting that the observed dichroism in Ref.  2 is peculiar to 
films. To date, the thin-film measurement of Ref. 2 is the only ARPES 
experiment in the literature reporting a non-zero dichroic effect at 
low-T.

We emphasize that a 1\% dichroic effect can be induced by an in-plane 
movement of O-atoms in the BiO layer by only 0.03 \AA, which is well 
below the accuracy with which lateral positions of surface 
atoms can be determined currently via surface sensitive probes. 
Given the sensitivity of dichroism to structrural details 
it will be prudent to keep the following further points in mind in 
interpreting dichroic effects and their temperature/doping dependencies: 
(1) ARPES is a surface-sensitive probe, and the surface structure of 
thin-films or single crystals will in general be different from bulk. We 
employed an ideal Bi-O terminated semi-infinite crystal in our 
computations because the detailed surface structure is unknown (x-ray 
scattering is not sensitive to surface structure); (2) Surface structure 
will likely be more sensitive than bulk to temperature; (3) Ref. 2 reports 
sample to sample variation in dichroism, which they ascribe to changes in 
domain structure. If so, such domain variations may become exaggerated in 
the surface region. Our simulations show that 1\% dichroic effect can 
result from only a 4\% change in the population of one twin; (4) 
Additional complications involve effects of disorder and layer-to-layer 
variations in structure in the surface region.

We would expect the computed momentum and energy dependencies of the 
dichroic signal to be even more sensitive to structural details than the 
size of the dichroic signal. Therefore, the detailed comparisons of the 
sort alluded to in the Comment between our computations on an ideal Bi-O 
terminated bulk structure and experiments on thin films, where the 
computations and measurements refer to different structures, are not of 
much relevance.

Finally, we note that structural issues with the study of Ref. 2 have been 
raised earlier in the comments contained in Refs. 4 and 5. We refer to 
those comments and responses for details of these and other issues raised 
previously in the literature on the study of Ref. 2.

We conclude that arguments of Norman et al. in their present Comment do 
not provide a significant basis for their claim that the geometric 
mechanism for explaining the observations reported in Ref. 2 is not 
viable. More generally, our study [1] highlights the importance of 
assessing structural issues before invoking exotic mechanisms for 
explaining unusual spectroscopic observations, especially in complex materials.

Work supported by the Division of Materials Science and Engineering  
(U.S.D.O.E.) under 07ER46352.

\end{document}